\documentclass[sigconf,nonacm]{acmart}
\usepackage{tabularx}
\usepackage{multirow}
\usepackage{booktabs} 
\usepackage{array} 
\usepackage{colortbl} 
\usepackage{graphicx}
\usepackage{tabularray}
\usepackage{soul} 
\usepackage{bm}
\usepackage{listings} 

\usepackage[frozencache=true,cachedir=minted-cache]{minted} 
\usepackage{xspace}
\usepackage{subcaption}
\newcommand*{\eg}{\textit{e.g.}\@\xspace}
\newcommand*{\ie}{\textit{i.e.}\@\xspace}

\newcommand*{\vs}{\textit{vs.}\@\xspace}
\newcommand{\sysname}{EyeQ\xspace}

\AtBeginDocument{%
  }

\begin{document}

\title{\sysname{}: Exploring Corneal Reflection for Mobile Context Sensing}

\author{Jaewook Lee}
\affiliation{%
  \institution{University of Washington}
  \city{Seattle}
  \state{Washington}
  \country{USA}
}

\author{Judith Amores}
\affiliation{%
  \institution{Microsoft Research}
  \city{Cambridge}
  \state{Massachusetts}
  \country{USA}
}

\author{Donghoon Shin}
\affiliation{%
  \institution{University of Washington}
  \city{Seattle}
  \state{Washington}
  \country{USA}
}

\author{Bo-Eun Choi}
\affiliation{%
  \institution{University of Washington}
  \city{Seattle}
  \state{Washington}
  \country{USA}
}

\author{Jason Kim}
\affiliation{%
  \institution{University of Washington}
  \city{Seattle}
  \state{Washington}
  \country{USA}
}

\author{Davin Win Kyi}
\affiliation{%
  \institution{University of Washington}
  \city{Seattle}
  \state{Washington}
  \country{USA}
}

\author{Seok-Young Kim}
\affiliation{%
  \institution{KAIST}
  \city{Daejeon}
  \country{Republic of Korea}
}

\author{Jon E. Froehlich}
\affiliation{%
  \institution{University of Washington}
  \city{Seattle}
  \state{Washington}
  \country{USA}
}

\author{Andrew D Wilson}
\affiliation{%
  \institution{Microsoft Research}
  \city{Redmond}
  \state{Washington}
  \country{USA}
}

\renewcommand{\shortauthors}{Lee et al.}

\begin{abstract}
What if a single face-facing camera could sense your surroundings---through your eyes? We introduce \textit{\sysname}, which explores corneal reflection as a modality for context-aware mobile interaction. Traditional vision-based context sensing relies on world-facing cameras that demand explicit aiming or dual-camera setups; \sysname instead shows that an off-the-shelf VLM can read scene context from corneal reflections captured with a smartphone camera. To recover usable detail from mobile corneal crops, \sysname pairs on-device iris-guided capture with server-side AI enhancement, then leverages VLMs to extract scene types, names, and objects from warped, unfamiliar eye-image sequences. In an $N{=}10$ study, we characterize how capture conditions shape reflection fidelity, evaluate what VLMs detect, misclassify, and hallucinate, and demonstrate interaction opportunities, including detecting a steering wheel to distinguish drivers from passengers. This work reframes the eye in HCI---from a target for gaze estimation to a passive mirror of one's surroundings---and examines the privacy tensions it raises.
\end{abstract}



\begin{teaserfigure}
\centering
  \includegraphics[width=.96\textwidth]{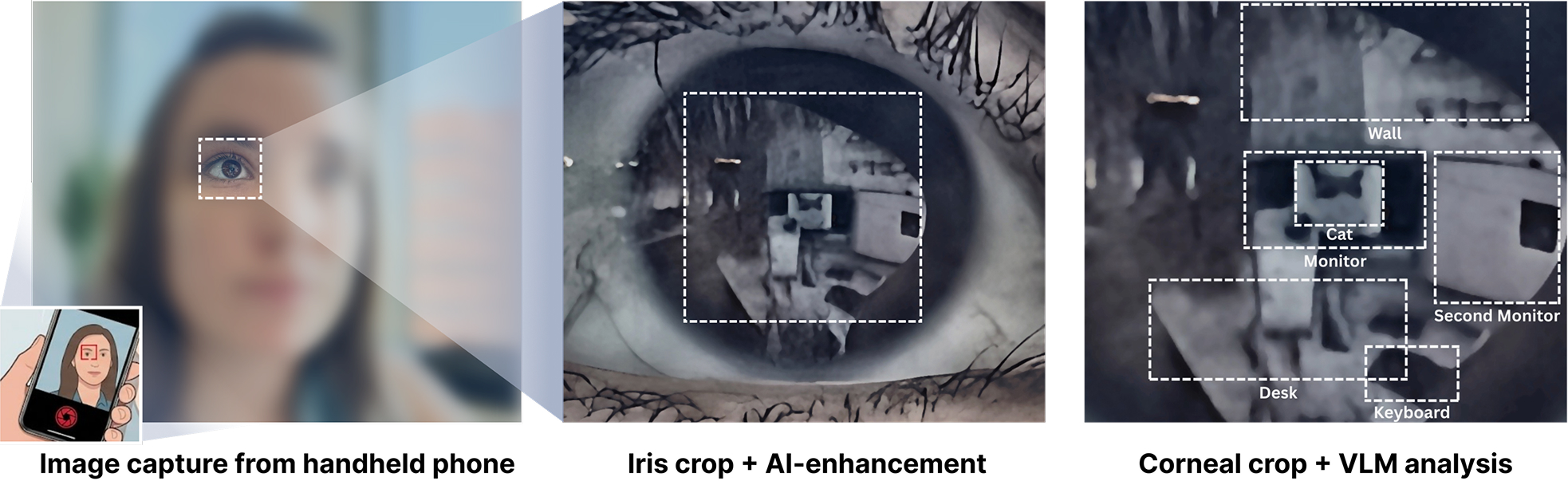}
  \caption{We introduce \textit{\sysname}, which passively extracts environmental context by capturing the user's cornea with a smartphone camera. \sysname employs a multi-stage pipeline---including eye detection, adaptive focus, auto-cropping, AI-based image restoration, and VLM analysis---to infer scene type, name, and visible objects from corneal reflection time-series. In this office scene, \sysname identifies two monitors, a keyboard, desk, and even a cat image displayed on screen.}
  \label{fig:teaser}
\end{teaserfigure}


\maketitle

\section{Introduction}
Context-aware mobile systems---from voice assistants that adapt to their surroundings~\cite{mayer2020worldgaze, lee2024gazepointar} to AR interfaces that recognize nearby objects~\cite{dogan2024xrobjects, lee2025sensibleagent}---depend on knowing what is around the user, yet acquiring this information remains cumbersome. Traditional vision-based context sensing systems require a world-facing camera to observe the scene and often a face-facing camera for gaze or attention~\cite{mayer2020worldgaze, lee2024gazepointar, lander2017heyebrid}, demanding active device aiming, raising social acceptability and privacy concerns~\cite{rico2010usable, alasakar2023investigating}, and excluding many everyday configurations where the only useful camera is oriented toward the user (\eg, phones held at a natural downward angle, laptops, webcams, propped-up tablets). A rich signal, however, is already visible to any single face-facing camera: the human cornea acts as a convex mirror, reflecting a wide-angle view of the surrounding scene~\cite{nishino2004worldineye} and effectively turning the user's eye into a passive scene camera.

\textit{Corneal reflections}---the specular reflections of the environment visible on the corneal surface---have been studied in computer vision~\cite{nitschke2013cornealimagingrevisited, nitschke2018cornealimaging}, where prior work has recovered environment maps and 3D scene geometry from eye images~\cite{nishino2004worldineye, nishino2006corneal, alzayer2024seeingtheworld}, and in security as a side-channel threat~\cite{backes2008compromisingreflections, backes2009tempestinateapot}. However, they remain largely unexplored as a sensing modality in Human-Computer Interaction (HCI). A key barrier is that existing pipelines assume controlled capture (\eg, DSLR optics, studio lighting) and often require substantial processing time, limiting their use on everyday devices. Within HCI, eye reflections have been used for near-device interaction, such as sensing grip posture and finger pose from screen glints~\cite{zhang2022reflectouch, nakamura2026reflectrace} or detecting input around reflective eyewear~\cite{grubert2016glasshands}, but not yet for broader scene understanding. Most relevant to our work, \textit{hEYEbrid}~\cite{lander2017heyebrid} uses a multi-camera wearable for calibration-free gaze estimation, but it treats the corneal image as a geometric cue for gaze direction rather than a source of scene information. We instead explore whether a vision-language model (VLM) can read scene context from low-resolution corneal reflections captured with a smartphone camera, while also confronting the privacy implications of this intimate signal.

To study this, we built \textit{\sysname}, an end-to-end pipeline that extracts scene information---scene type (\ie, indoor \vs outdoor), scene name (\eg, kitchen), and object inventories---from corneal reflections captured by a smartphone camera over time. The core challenge is that a corneal reflection occupies only a tiny fraction of the face image, yielding eye crops that are often too small, blurry, and dark for direct analysis. \sysname addresses this through a split architecture. On-device, it keeps each frame sharp, well-exposed, and centered on the cornea through four stages: gyroscope-based motion gating, \textit{MediaPipe Iris}~\cite{mediapipe} for steering auto-exposure and auto-focus to the eye, 48-megapixel (MP) photo capture (rear camera as a stand-in\footnote{We capture with the rear camera to test corneal sensing at the upper end of current smartphone optics. We pair it with a mirrored display to simulate a future phone with a comparable front camera. That gap is closing---front-camera resolution has climbed steadily~\cite{counterpoint2021}, and some Android phones now ship with 50\,MP front sensors~\cite{gsm-moto, gsm-oppo}.}), and iris-guided corneal cropping. Server-side, \sysname recovers faint scene detail from the raw crop in five stages: \textit{Real-ESRGAN}~\cite{wang2021realesrgan} 4$\times$ super-resolution, CLAHE local contrast enhancement, \textit{NAFNet}~\cite{chen2022nafnet} denoising, unsharp masking, and saturation boosting. Both raw and enhanced versions are batched over time and sent to three VLMs---Google \textit{Gemini 3.1 Pro}~\cite{gemini31pro}, \textit{Gemini 3 Flash}~\cite{gemini3flash}, and \textit{Gemini 3.1 Flash Lite}~\cite{gemini31flashlite}---for progressive scene analysis. To our knowledge, this is the first work to infer scene content from corneal reflections captured with smartphone-grade optics, and the first to apply VLMs to these low-resolution signals.

To evaluate feasibility, we conducted an $N{=}10$ within-subjects study that varied capture conditions (\eg, motion, phone-to-eye distance, angle, and lighting) and examined physiological factors (\eg, eye color) across indoor and outdoor scenarios, measuring their impact on image quality and VLM scene recognition. Results show that VLMs can reliably classify indoor \vs outdoor scenes (96.4\%) and detect objects from corneal reflections, including monitors (87.5\% of office trials), windows (100\% in kitchen), trees (40\% outdoors), and even a cat image on a screen. Most false positives were misclassifications of real optical evidence rather than hallucinations, suggesting VLMs see more in the cornea than they correctly name. To close that gap, we asked whether corneal reflections could be combined with other context available to a phone, such as GPS, time, motion, or the rest of the frame around the eye. Providing a minimal scene prior (\eg, ``the user is in a kitchen'') improved detection across all three scenes and halved false positives. We complement the study with five demonstrations of interactions this signal can support, from warning a phone-distracted pedestrian about an approaching vehicle to pausing a voice assistant when someone else is nearby. Finally, we discuss how future advances could extend this capability, alongside the ethical tensions of corneal reflection, from participants' reactions to seeing their own eyes, to our decision not to fine-tune models or release our 10,635-image dataset.

In summary, our contributions include:
\begin{enumerate}
    \item \textit{\sysname}, a pipeline that infers scene information from corneal reflections captured with smartphone-grade optics, combining on-device iris-guided capture with server-side AI enhancement and VLM scene analysis.
    \item Findings from an $N{=}10$ study characterizing what off-the-shelf VLMs detect, misclassify, and hallucinate, how capture conditions shape reflection quality, and how participants perceive comfort and privacy implications---establishing a zero-shot benchmark.
    \item Proof-by-demonstration scenarios illustrating interaction opportunities and risks this capability raises.
\end{enumerate}

\section{Related Work}
Corneal reflections have been studied across computer vision, privacy, and HCI. We build on this interdisciplinary foundation to examine the potential and risks of analyzing eye reflections, as well as how \sysname extends this space by enabling such techniques with smartphone-grade optics and VLMs.

\subsection{Scene Understanding from Eye Reflections}
The human cornea acts as a convex mirror, providing a natural reflective surface for environment capture. Seminal work by Nishino \& Nayar modeled the eye as a catadioptric imaging system, recovering wide-angle environment maps from single eye images~\cite{nishino2004worldineye, nishino2006corneal}. They showed that corneal reflections encode both panoramic views of the surroundings and retinal images aligned with gaze, laying the foundation for using eyes as scene cameras. Later studies refined this mapping with more accurate, aspherical eye models~\cite{nakazawa2016registration}.

A complementary thread treats the cornea as a light probe for estimating scene illumination. Tsumura \textit{et al.} estimated the directions of multiple light sources from eye reflections~\cite{tsumura2003estimating}. Nishino \& Nayar then extracted dense spherical illumination maps for face relighting~\cite{nishino2004relighting}, and Wang \textit{et al.} separated specular reflections from iris texture to isolate light sources across varied eye colors~\cite{wang2005separating}. Corneal illumination estimates have also supported downstream tasks, including face~\cite{nishino2005facerecognition} and place~\cite{ohshima2021visual} recognition.

Subsequent work improved corneal image quality and extended it to 3D reconstruction. Nitschke \& Nakazawa applied super-resolution to eye videos by aggregating reflections over time~\cite{nitschke2012super}. Nishino \& Nayar modeled stereo geometry between both eyes for triangulating reflected scene points~\cite{nishino2004worldineye, nishino2006corneal}, with related work recovering scene geometry from corneal reflections of known screen patterns~\cite{nitschke2009displaycalib, schnieders2010displayrecon, nitschke2011display}. More recently, Liang~\textit{et al.} recovered lighting, shape, and appearance from a single unconstrained eye image via inverse rendering~\cite{liang2024eyeir}, while Alzayer~\textit{et al.} jointly estimated 3D eye poses and a NeRF from short face videos, treating each corneal image as a novel view for scene reconstruction~\cite{alzayer2024seeingtheworld}.

While we are inspired by this body of work, it typically relies on DSLR-quality images under controlled conditions (\eg, tripod-mounted cameras, studio lighting). Instead, we explore whether corneal reflection cues can be captured by a handheld smartphone and analyzed by VLMs for context-aware HCI applications.

\subsection{Gaze and Attention via Corneal Reflections}
Corneal reflections are foundational to model-based gaze tracking, though the term refers to a different signal than in our work. Classic pupil-center corneal reflection (PCCR) methods use \textit{glints}---point reflections from known infrared sources---as geometric markers, estimating gaze from their position relative to the pupil~\cite{guestrin2006general,hansen2009eye}. \sysname instead analyzes reflected \textit{scene content}, not glint geometry.

A closer line of work estimates gaze by analyzing scene imagery visible on the corneal surface. Nakazawa~\textit{et al.} estimated a 3D point of gaze using coded IR patterns~\cite{nakazawa2012point}, while Nitschke~\textit{et al.} estimated point-of-gaze directly from the corneal image~\cite{nitschke2013see}. Related approaches leverage aspherical eye models for geometric gaze estimation~\cite{nagamatsu2010gaze}, estimate point-of-regard from corneal images~\cite{takemura2014estimating}, and enable calibration-free view estimation on mobile head-worn cameras~\cite{nakazawa2015mobile}. Notably, Takemura~\textit{et al.} found the object of focus from its corneal reflection~\cite{takemura2014estimation}, while \textit{HiFiGaze}~\cite{kim2026hifigaze} shows that screen reflections visible on the cornea can refine mobile gaze estimation.

\subsection{HCI Systems Using Corneal Reflections}
Several HCI systems have leveraged corneal reflections for context-aware interaction~\cite{nitschke2013cornealimagingrevisited}. For example, \textit{ReflecTouch}~\cite{zhang2022reflectouch} and \textit{ReflecTrace}~\cite{nakamura2026reflectrace} analyze finger occlusions in the corneal reflection of the screen to infer grip posture and mid-air hover gestures, respectively. Most relevant to our work, \textit{hEYEbrid}~\cite{lander2017heyebrid} mounted two eye-facing cameras---one infrared, one for corneal imaging---on a glasses frame for calibration-free gaze estimation, using corneal reflection as a geometric cue for gaze direction. Beyond personal devices, Nagamatsu \textit{et al.} demonstrated gaze-responsive public displays using corneal reflections without calibration~\cite{nagamatsu2014publicuse}. Recent work has also curated datasets pairing corneal reflections with scene views~\cite{ohshima2021visual,du2024cred}, though neither is publicly available due to privacy concerns.

HCI researchers have also explored external reflective surfaces beyond the cornea. \textit{GlassHands}~\cite{grubert2016glasshands} demonstrated around-device input by analyzing reflections of the phone in eyewear such as sunglasses and visors, an approach later adapted to use corneal reflections directly~\cite{schneider2017arounddevice}. Other work has captured reflections on physical mirrors~\cite{plasencia2014throughglass,lee2017reflector,zhou2024reflectedreality} and device-mounted mirrors~\cite{wong2016backmirror} to extend interaction areas and sense occluded objects.

Prior work uses corneal reflections primarily for near-device interaction and gaze estimation; we instead investigate them as a source of environmental context, leveraging VLMs to extract scene information from low-resolution corneal imagery. Our approach also requires only a single smartphone camera, simplifying the hardware demands of prior eye-driven context systems (\eg, \textit{3C-hEYEbrid}~\cite{lander2017heyebrid}, \textit{WorldGaze}~\cite{mayer2020worldgaze} and \textit{GazePointAR}~\cite{lee2024gazepointar}, all of which require both eye- and world-facing cameras) and removing the need to aim the device at the scene.

\subsection{Privacy Concerns around Reflections}
A growing body of work has shown that reflections in everyday objects can inadvertently leak sensitive visual information. High-resolution images of corneal reflections can expose bystanders, screen contents, or entire environments~\cite{backes2008compromisingreflections, backes2009tempestinateapot, jenkins2013identifiablebystander}. Prior work has also reconstructed typed input from reflections in sunglasses~\cite{raguram2011ispy}, recovered screen contents via teapots, spoons, plastic bottles, and eyeglasses~\cite{backes2008compromisingreflections}, exploited multi-bounce reflections to reconstruct obscured input~\cite{xu2013seeingdouble}, and even identified websites browsed during video calls using eyeglass glints~\cite{long2023privateeye}. Recent advances in computer vision push these threats further, with neural rendering techniques extracting 3D scene geometry from eye reflections~\cite{alzayer2024seeingtheworld}. These studies collectively demonstrate that visual side-channels pose a credible privacy risk in both public and private settings. As corneal reflection sensing becomes feasible with smartphone-grade optics, we examine both its potential and privacy risks.

\section{E\lowercase{ye}Q Implementation}
\label{sec:system}
\begin{figure*}[hbt!]
  \centering
  \includegraphics[width=\linewidth]{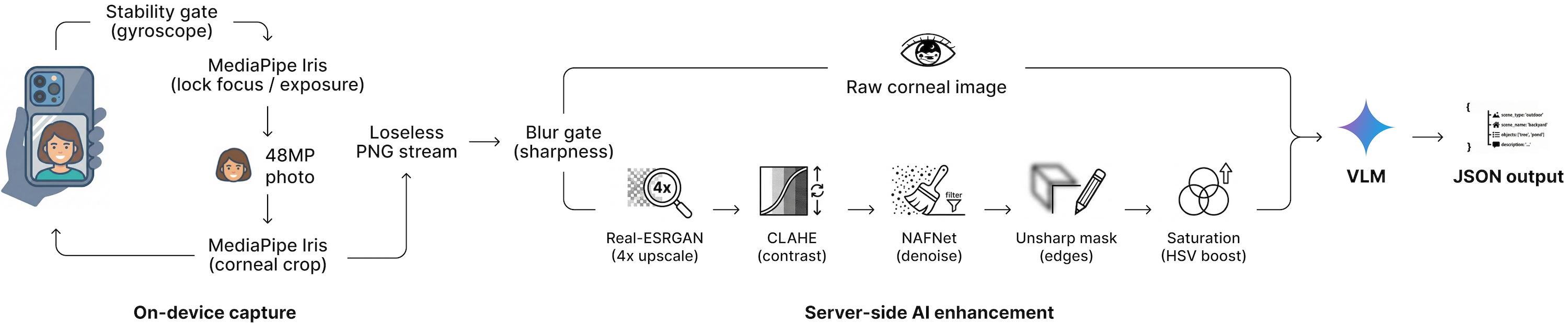}
  \caption{\textbf{\sysname pipeline overview.} An iOS client captures iris-targeted 48\,MP photos, streams corneal crops to a server for AI-based enhancement, and sends raw--enhanced pairs to three VLMs in parallel for progressive scene analysis.}
  \vspace{-10px}
  \label{fig:system-diagram}
\end{figure*}

To explore corneal reflection sensing with smartphone-grade optics, \textit{\sysname} pairs on-device capture with server-side enhancement in a single pipeline (\autoref{fig:system-diagram}). This split architecture is needed because the cornea occupies a tiny fraction of the face image, yet vision interpretation demands more detail than the raw crop provides. \sysname captures iris-targeted 48\,MP photos, streams each corneal crop to a GPU server for AI-based restoration, and sends raw--enhanced pairs to VLMs for analysis over time.

\subsection{Mobile Capture Client}
\label{sec:client}
We developed an iOS client targeting an iPhone 16 Pro's 48\,MP rear
sensor (8064$\times$6048, 24\,mm, $f$/1.8). We chose the rear sensor to test corneal sensing at the upper end of current smartphone optics. The front camera on our device captures at 12\,MP, which our piloting found insufficient for corneal analysis. To give participants a sense that they were still using a smartphone, we pair the rear camera with a mirrored display so they have a screen to look at (see \autoref{fig:apparatus} inset). We drew inspiration from how people already use external preview displays or a foldable phone's outer screen to take selfies with the rear camera. Moreover, front cameras are catching up quickly~\cite{counterpoint2021}---the iPhone 17 Pro now has an 18\,MP front sensor and 50\,MP front sensors have begun shipping on Android phones~\cite{gsm-moto, gsm-oppo}---hence the rear camera serves as a reasonable stand-in. The client handles three tasks: real-time iris localization, high-resolution photo capture, and corneal cropping and transmission.

\textbf{Real-time iris localization.} Because the camera's default auto-exposure and auto-focus (AE/AF) optimize for the face, we first have to steer capture toward the cornea. We run Google's \textit{MediaPipe Iris}~\cite{mediapipe, mediapipe_prebuilt} on each live preview frame (1920$\times$1080 at 60\,FPS), choosing it over classical eye detectors (\eg, Haar cascades) that lose reliability at the upward viewing angles typical of handheld phone use. MediaPipe Iris detects 10 iris landmarks---five per eye, comprising a center point and four cardinal boundary points---from which we select the larger iris and compute its centroid. Before each capture, \sysname locks the camera's AE/AF point of interest onto the detected iris center, ensuring the photo is optimized for the corneal region. \sysname also continuously estimates phone-to-eye distance using a pinhole camera model and the known human iris diameter ($\approx$11.7\,mm~\cite{bergmanson2017size, rufer2005white, baumeister2004comparison, hashemi2015white}), warning the user when the phone is closer than the lens's minimum focus distance ($\sim$15\,cm).

\textbf{High-resolution photo capture.} Maximizing pixel density on the cornea is critical. We empirically evaluated five iOS camera configurations---48\,MP ProRAW, 48\,MP HEIC, 24\,MP Deep Fusion, 12\,MP burst mode, and 2--5$\times$ digital zoom---finding 48\,MP HEIC at 1$\times$ zoom with quality prioritization and flash disabled offered the best trade-off. ProRAW forgoes Apple's computational photography pipeline (\ie, Deep Fusion and Smart HDR), which proved critical for corneal detail; burst mode sacrifices resolution; and digital zoom amplified noise and blur. Before each capture, \sysname waits for the iris-steered AE/AF to stabilize, and a gyroscope-based motion gate suppresses capture during rapid hand movement (rotation rate $>$\,0.4\,rad/s). This loop repeats as fast as the hardware allows---a ``\textit{lucky imaging}''~\cite{fried1978probability, law2006lucky} strategy at $\sim$0.5\,Hz, capturing many frames in hopes of obtaining at least one sharp corneal image.

\textbf{Corneal crop and transmission.} After each 48\,MP capture, \sysname runs a second MediaPipe Iris pass on the embedded preview thumbnail to quickly obtain landmark positions. These normalized coordinates are mapped to the full-resolution image with aspect-ratio correction. \sysname then crops a square region around the larger iris with 15\,px padding from the full photo. Because the 48\,MP sensor stores pixels in Display P3 with 10-bit depth, we convert the crop to 8-bit sRGB and encode it as a lossless PNG to preserve fine specular detail. Each crop is streamed to the server over a persistent WebSocket connection.

\subsection{Server-Side Enhancement}
Even at 48\,MP, the corneal crop typically spans only 150--300\,px---which is often too small for reliable VLM interpretation. The backend receives each corneal image and applies fast, multi-stage AI restoration to recover faint scene detail.

\textbf{Quality gating.} Not all incoming crops are worth processing, as motion blur, blinks, and inaccurate iris detections produce unusable frames. We compute a sharpness score from Sobel gradient variances over the center 50\% of the crop and reject frames below an empirically set threshold ($S < 100$).

\textbf{Enhancement stages.} Accepted crops pass through five stages, each with parameters that adapt to the input quality:
\begin{enumerate}
    \item \textit{Real-ESRGAN 4$\times$}~\cite{wang2021realesrgan} super-resolution upscales the crop to 600--1200\,px. We chose this over face-restoration models (\eg, \textit{GFPGAN}~\cite{wang2021gfpgan}), which impose facial geometry priors that distort scene reflections, and over diffusion-based approaches, which are slower and risk hallucinating detail.
    \item \textit{Contrast-Limited Adaptive Histogram Equalization (CLAHE)} on the L channel of LAB color space boosts local contrast, pulling detail from dark iris regions (clip limit scales with input brightness: 1.0--1.5).
    \item \textit{NAFNet}~\cite{chen2022nafnet} denoising suppresses noise amplified by CLAHE while preserving fine edges; we found it retained specular detail better than Gaussian smoothing or bilateral filter.
    \item \textit{Unsharp Masking} in float32 (to avoid clipping corneal highlights near 255) re-sharpens structural edges such as reflection contours (strength 1.0--1.5 based on input sharpness).
    \item \textit{Saturation Boost} in HSV space increases color separation (factor ranges 1.05--1.15 based on input contrast).
\end{enumerate}
All thresholds were empirically tuned. Because these enhancement models are trained on everyday photos rather than corneal reflections, they can reveal detail not apparent in the raw crop but may also introduce artifacts. \sysname therefore stores both raw and enhanced versions for VLM analysis. Per-component latencies are reported below in Table~\ref{tab:latency}.

\begin{table}[h]
\centering
\caption{Per-component latency ($M \pm SD$, in seconds). Server-side measurements on an NVIDIA RTX 4080.}
\scriptsize
\label{tab:latency}
\begin{tabular}{l r @{\hskip 1.5em} l r}
\toprule
\multicolumn{2}{c}{\textbf{iOS Capture}} & \multicolumn{2}{c}{\textbf{Server-Side AI Enhancement}} \\
\cmidrule(r){1-2} \cmidrule(l){3-4}
MediaPipe Iris~\cite{mediapipe} & $0.008 \pm 0.005$ & Real-ESRGAN 4$\times$~\cite{wang2021realesrgan} & $0.10 \pm 0.05$ \\
Lock Auto-Focus/Exposure & $0.39 \pm 0.23$ & CLAHE & $0.003 \pm 0.001$ \\
48MP Photo Capture & $1.29 \pm 0.22$ & NAFNet~\cite{chen2022nafnet} & $0.31 \pm 0.05$ \\
Corneal Crop & $0.05 \pm 0.01$ & Unsharp Mask & $0.02 \pm 0.006$ \\
Image Stream (iOS $\to$ Server) & $0.10 \pm 0.12$ & Saturation Boost & $0.008 \pm 0.002$ \\
\bottomrule
\addlinespace[0.5em]
\toprule
\multicolumn{4}{c}{\textbf{VLM Scene Understanding + LLM Judge}} \\
\cmidrule{1-4}
Gemini 3.1 Pro~\cite{gemini31pro} & $31.88 \pm 24.24$ & Gemini 3 Flash~\cite{gemini3flash} & $10.61 \pm 2.05$ \\
Gemini 3.1 Flash Lite~\cite{gemini31flashlite} & $8.48 \pm 1.42$ & LLM Judge (GPT-4.1~\cite{gpt41}) & $1.06 \pm 0.24$ \\
\bottomrule
\end{tabular}
\end{table}

\subsection{Scene Analysis}
\label{sec:scene-analysis}
Once corneal crops have been captured and enhanced, \sysname sends them to three VLMs in parallel---Google \textit{Gemini 3.1 Pro}~\cite{gemini31pro}, \textit{Gemini 3 Flash}~\cite{gemini3flash}, and \textit{Gemini 3.1 Flash Lite}~\cite{gemini31flashlite}---all with the \textit{highest} available thinking budget, each with a different performance--latency trade-off. Images are grouped into windows and sent as raw--enhanced pairs so each VLM can leverage whichever is clearer. Each VLM progressively builds a separate scene understanding, dispatching a new window once the previous call returns and at least four new image pairs are available: every call receives the current master scene state and returns a refined version, accumulating \texttt{confirmed} objects (clearly identifiable) and \texttt{tentative} objects (educated guesses) while pruning false positives. The output follows a fixed JSON schema containing \texttt{scene\_type} (\ie, indoor \textit{vs.}\ outdoor), \texttt{scene\_name} (\eg, office, kitchen), \texttt{confirmed\_objects} and \texttt{tentative\_objects} lists, and a brief \texttt{scene\_description}. Responses are used as they arrive, so faster models can support interaction sooner (see \S\ref{sec:proof-by-demonstration}) while slower ones contribute more reliable scene state, a tiering strategy inspired by \textit{WorldScribe}~\cite{chang2024worldscribe}.

Because VLM outputs use open-ended vocabulary (\eg, ``table lamp'' \vs ``desk lamp''), \sysname applies an OpenAI \textit{GPT-4.1}~\cite{gpt41} LLM judge to normalize detected objects and resolve synonyms, making downstream matching more robust. The full VLM and LLM judge prompts are provided in Appendices~\ref{sec:vlm-prompt} and~\ref{sec:llm-prompt}.

\section{User Study}
\label{sec:user-study}
To assess technical feasibility, we conducted an $N{=}10$ within-subjects study using \sysname as the capture instrument. The primary goal was to characterize how capture conditions (\eg lighting, phone-to-eye distance, angle, motion) and physiological factors (\eg eye color) affect image quality and VLM scene analysis across indoor and outdoor scenarios. We also gauged comfort and privacy perceptions around corneal sensing before and after participants saw their own eye images. Each session comprised a pre-study questionnaire, a tutorial, 16 capture trials, and a post-study debrief.

\begin{table}[h]
\caption{Participant demographics (N=10)}
\scriptsize
\begin{tabular}{cccccc}
\toprule
\textbf{PID} & \textbf{Gender} & \textbf{Age} & \textbf{Eye color} & \textbf{Contacts?} & \textbf{Consent to image inclusion?}\\
\midrule
P1 & Male & 26 & Brown & No & No \\
P2 & Female & 30 & Brown & No & Yes \\
P3 & Male & 24 & Dark brown & No & No \\
P4 & Male & 22 & Dark brown & No & No \\
P5 & Female & 36 & Green & Yes & Yes \\
P6 & Male & 22 & Green & No & Yes \\
P7 & Female & 20 & Hazel & No & Yes \\
P8 & Male & 30 & Green & No & Yes \\
P9 & Female & 31 & Brown & Yes & Yes \\
P10 & Female & 27 & Blue & No & No \\
\bottomrule
\end{tabular}
\label{tab:demographics}
\end{table}

\subsection{Participants}
We recruited 10 adults ($M_{age}$=26.8; $SD_{age}$=5.0) through social media posts and snowball sampling. The sample included five female and five male; five had dark-colored eyes and five had light-colored eyes. Two wore contacts. See Table~\ref{tab:demographics}. We excluded participants who wore glasses, as frames occlude the cornea and lenses add competing reflections. This was a confound to control for, not a limit of the approach, since prior work has used eyewear reflections~\cite{grubert2016glasshands}. We also excluded those with visual, auditory, or mobility impairments. Demographics were collected via a pre-study screening questionnaire. We aimed to balance eye-color representation across the global distribution (79\% brown, 8--10\% blue, 5\% hazel, 2\% green~\cite{dorgaleleh2020molecular}), though this limited our sample size, as light-eyed participants were harder to recruit.

\subsection{Apparatus}
\label{sec:apparatus}
\begin{figure*}[hbt!]
  \centering
  \includegraphics[width=\linewidth]{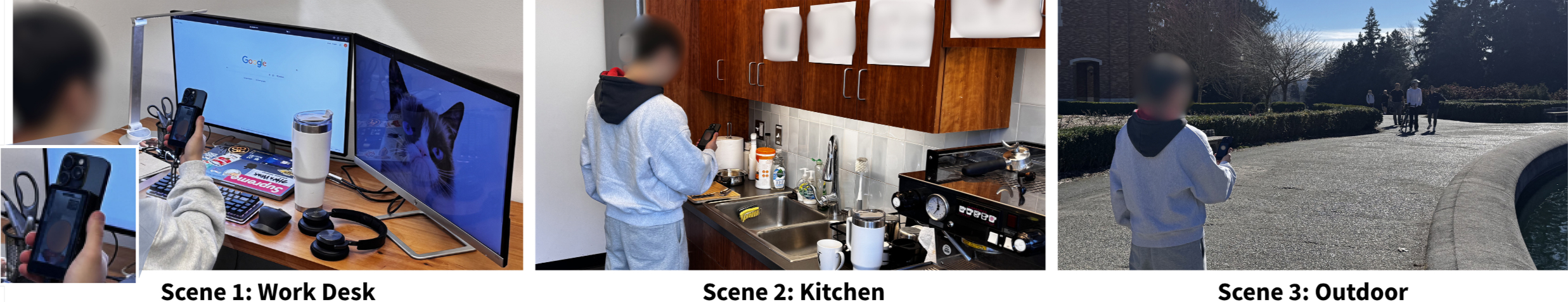}
  \vspace{-0.5cm}
  \caption{Study apparatus. The three scenarios (office, kitchen, campus), their ground-truth objects, and the rear camera phone-holding posture with a magnetic mirrored display (bottom-left inset).}
  \vspace{-0.3cm}
  \label{fig:apparatus}
\end{figure*}

We gave participants an iPhone 16 Pro running \sysname, using its rear camera to capture 48\,MP photos and a magnetically mounted mirrored display to preserve the form factor of an ordinary phone (\autoref{fig:apparatus}). Phone-to-eye distance was measured with a flexible tape measure and phone angle with a digital protractor. For tripod trials, the phone was fixed using a gooseneck holder; for handheld trials, both \sysname and the researcher monitored posture drift, warning participants of substantial deviation. Lighting was measured as incident illuminance on the face using a light-meter app. Unlike prior corneal reflection work that uses controlled studio lighting, we relied only on typical ambient lighting (\eg, ceiling fixtures), and scenes were populated with everyday objects.

\subsection{Procedure}
Each session consisted of four phases: (1) pre-study attitude questions ($\sim$10\,min), (2) tutorial ($\sim$10\,min), (3) capture trials ($\sim$90\,min), and (4) post-study debrief ($\sim$10\,min), for a total of about two hours.

\textbf{Pre-study questionnaire.}
Before introducing capture tasks, we asked four five-point Likert questions probing: (1) comfort with visible camera use in public, (2) prior awareness that eye reflections reveal surroundings, (3) comfort with a hypothetical ``eye-zoom'' capability on everyday phones, and (4) comfort if eye data were kept on-device. Participants were asked to explain each rating.

\textbf{Tutorial.}
We demonstrated the phone-holding orientation used in the experiment, then ran a short practice trial where participants looked at objects on an office desk (scenario 1) for $\sim$1 minute.

\textbf{Controlled capture trials.}
Each participant completed 16 trials across three scenarios, varying four two-level factors: (1) \textit{Motion}: tripod \vs handheld; (2) \textit{Distance}: $\sim$20\,cm \vs $\sim$40\,cm phone-to-eye distance; (3) \textit{Angle}: $\sim$30$^\circ$ \vs $\sim$60$^\circ$ below eye level; and (4) \textit{Lighting}: high (floor lamp added) \vs low (lamps off, ceiling dimmed to 20\%). Distance and angle values were informed by prior work on phone-holding posture~\cite{bababekova2011font, yoshimura2017smartphone, lee2015head} and refined through piloting. The 40\,cm, 60$^\circ$ condition approximates typical phone use; the closer levels are less common but remain within that reported range, and also preview what a higher-resolution sensor might one day capture at the farther distance.

To keep the session manageable, not all factors were crossed in every scenario. The \textit{Office} (10 trials) fully crossed motion, distance, and angle ($2 \times 2 \times 2 = 8$) at typical office illumination ($M=402$ lux; $SD=124$), plus two lighting-isolation trials ($M_{high}=588$ lux, $SD_{high}=179$; $M_{low}=101$, $SD_{low}=27$) at a fixed posture (20\,cm, 30$^\circ$, tripod). The \textit{Kitchen} (5 trials; $M=424$ lux, $SD=154$) crossed distance and angle ($2 \times 2 = 4$), all handheld, plus one trial with \sysname propped on a counter stand, simulating propped-up tablets and laptop webcams. \textit{Campus} (1 trial; $M=9427$ lux, $SD=2661$) had no factorial control, under naturally varying ambient conditions. The scenarios progressively relaxed control: the office controlled sunlight (blinds closed) and body motion (seated); the kitchen introduced standing and walking; campus added uncontrolled lighting, sustained walking, and squinting in bright sun.

Each trial followed a fixed script: the researcher set the target distance and angle, then ran capture for 2 minutes (office), 3 minutes (kitchen), or 5 minutes (campus). Participants could look around freely but were asked to move a little more slowly than typical and pause briefly when fixating, as AE/AF and 48\,MP capture require settle time on current hardware. Breaks were offered between trials.

\textbf{Scenes and tasks.}
Each scenario contained 12 ground-truth objects against which we scored VLM outputs. The \textit{Office} included: desk, monitor, secondary monitor, image of a cat, mouse, keyboard, notebook, pen cup, desk lamp, coffee mug, eyeglasses, and headphones. The \textit{Kitchen} included: fridge, sink, faucet, coffee machine, coffee cup, bowl, fork, cutting board, pot, spatula, window, and a person (the researcher, periodically entering the scene). Here, we designed five short naturalistic tasks---opening the fridge, chopping on the cutting board, inspecting the sink, checking whether a cup is clean, and reading coffee-machine labels---so participants naturally varied proximity (\eg, standing upright \vs bending to inspect lower objects) and gaze direction. \textit{Campus} included: fountain, bench, trees, bushes, grass, fire hydrant, people, path, lampposts, trash can, road sign, and buildings. We included the campus scenario to probe whether current hardware can extract any scene information in a fully unconstrained setting.

\textbf{Post-study debrief interview.}
After the capture trials, we conducted a semi-structured interview mirroring the pre-study questions. We first asked about comfort with an app inferring surroundings from eye reflections, then showed participants their own corneal images and asked whether seeing the data changed their comfort level. We also showed a rear-camera photo of the same scene for comparison, asked whether the corneal images were clearer or blurrier than expected, what applications they envisioned, and their likelihood of adopting a future version of \sysname.

\subsection{Measures and Analysis}
\label{sec:analysis}
Because the study focuses on data collection rather than real-time interaction (which we demonstrate separately in \S\ref{sec:proof-by-demonstration}), only the raw corneal crops were saved during the trials; AI enhancement and VLM analysis were performed post-hoc. The raw--enhanced image pairs were then batched into 10-second sliding windows, simulating \sysname's progressive analysis, and run through all three VLMs (Gemini 3.1 Pro, Gemini 3 Flash, Gemini 3.1 Flash Lite) independently, each three times, to compare model accuracy across capture conditions. The LLM judge scored each trial's output against the scenario's 12-object ground truth, classifying detections as true positives (TPs), false positives (FPs), or false negatives (FNs), from which we computed precision, recall, and F1 score. Two researchers verified the judge's scoring, reviewing outputs for incorrectly assigned TPs, FPs, and FNs. We also analyzed Likert ratings and open-ended responses on comfort and privacy perceptions. The judge prompt and ground-truth definitions are in Appendix~\ref{sec:llm-prompt}.

The same two researchers then took a closer look at false positives, as an aggregate FP count does not distinguish whether the VLM invented something that was never there or saw something and named it wrong. They independently reviewed every FP against the corresponding raw and enhanced crops, labeling each as a \textit{misclassification} (optical evidence for a real object was present but mislabeled---\eg, a bright patch from an illuminated window labeled ``monitor'') or a \textit{hallucination} (no corresponding optical evidence in the image), resolving disagreements through discussion.

Finally, because corneal reflections captured on a smartphone are low-resolution, we tested whether fusing them with other context sources could reduce interpretation errors. We prepended a single sentence naming the scene to the VLM prompt (\eg, ``The user is in a kitchen''), which we call a ``\textit{minimal scene prior},'' and re-ran the full evaluation. This is not an artificial boost: it simulates what a phone can already assemble from sensors it has (excluding a world-facing camera), such as GPS, time, motion, activity state, ambient light, audio, or the rest of the frame around the eye crop.

\subsection{Ethics and Privacy}
An eye crop may not constitute personally identifiable information today (since this is not a retinal scan), but advancing AI could expose more from these images. Given this uncertainty, we took a conservative approach. Our IRB approved the study after an extended discussion reflecting these novel considerations; the approved protocol does not permit dataset release or use for model training. Participants could opt out of having their eye images appear in this paper; many did, so we also show corneal images from the research team. Recruitment was challenging, as several prospective participants declined once we explained that the study requires close-up eye imaging, citing concerns about data misuse. We discuss these concerns and potential mitigation strategies in \S\ref{sec:privacy-discussion}.

\section{Results}
We analyzed 10,635 corneal images across 10 participants (see \autoref{fig:participant-eyes}; additional examples in Appendix~\ref{sec:researcher-eyes}). We report five analyses: overall feasibility, capture condition effects, object-level detection and error patterns, a minimal scene-prior ablation, and participant perceptions. We evaluated three Gemini models (3.1 Pro, 3 Flash, and 3.1 Flash Lite; all \textit{high} thinking level), running each three times to account for variability. A Friedman test showed no significant effect of model on F1 score ($\chi^2(2, N{=}10) = 3.20$, $p = .20$). That said, Flash achieved the highest F1 and most detections, so we report Flash results below; per-model breakdowns are in \autoref{tab:scene-model}.

\subsection{Can VLMs Understand Corneal Reflections?}
Though likely never trained on corneal imagery, VLMs still extracted meaningful scene information from this low-resolution signal---albeit with limitations. All results below are zero-shot, with no fine-tuning or domain adaptation.

\begin{figure}[hbt!]
  \centering
  \includegraphics[width=.9\linewidth]{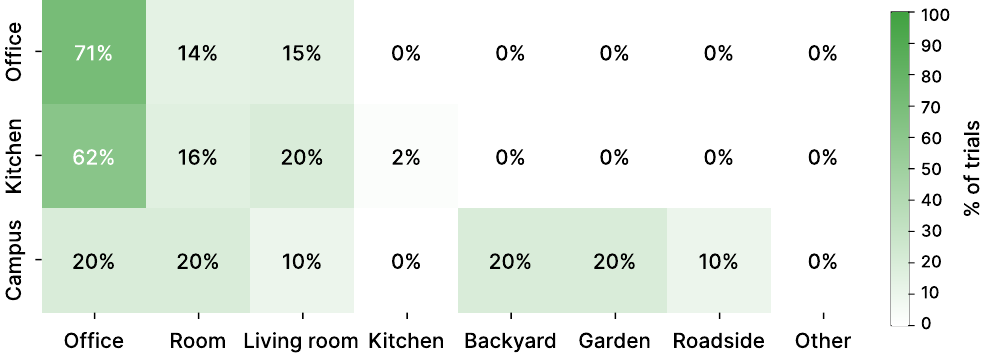}
  \caption{Scene name confusion matrix (Flash, without minimal scene prior). Indoor \textit{vs.}\ outdoor classification was reliable (96.4\%), but specific scene names were harder: office was correctly identified in 71\% of trials, kitchen was mislabeled ``office'' in 62\%, and campus received varied guesses (\eg, ``backyard,'' ``garden'').}
  \label{fig:scene-confusion}
\end{figure}

\textbf{VLMs classified scene type reliably but struggled with scene names.} Flash correctly classified scene type (indoor \textit{vs.} outdoor) in 96.4\% of trials (135/140; lighting trials analyzed separately).
However, predicting specific scene names proved harder: office was labeled correctly in 71\% of cases, but kitchen was mislabeled ``office'' in 62\% of trials, and campus received a mix of indoor (\eg, ``room'') and outdoor (\eg, ``backyard'', ``garden'') guesses (see~\autoref{fig:scene-confusion}).

\textbf{VLMs detected real objects across all three scenes.} A Friedman test revealed a significant effect of scene on F1 ($\chi^2(2, N{=}10) = 14.60$, $p < .001$), with office performing best ($M{=}.295$, $SD{=}.090$), followed by kitchen ($M{=}.157$, $SD{=}.033$) and campus ($M{=}.078$, $SD{=}.106$), all pairwise comparisons significant (Office--Kitchen: $p < .01$; Office--Campus: $p < .01$; Kitchen--Campus: $p < .05$). Still, \sysname detected real objects in every scene. In the office ($N{=}80$), Flash detected the monitor in 87.5\% of trials, the desk in 81.2\%, and the keyboard in 37.5\%. In the kitchen ($N{=}50$), the window was detected in 100.0\% of trials and a person in 18.0\%. Even outdoors on a campus ($N{=}10$), trees were identified in 40.0\% of trials, with buildings, people, and a road sign each detected at least once. These detections skew toward larger objects, and the small items in our ground truth (\eg, pen cup, fork, eyeglasses) went undetected, lowering F1. For now, this suggests corneal sensing is best suited for applications needing coarse context and large, high-contrast objects.

\begin{table*}[t]
\caption{VLM scene recognition performance by scene and model, without and with minimal scene prior. F1-score (F1), Precision (P), and Recall (R) reported as $M \pm SD$. TP = avg true positive count, FP = avg false positive count (misclassifications and hallucinations), FN = avg false negative count (missed), N = number of trials. Each scene contains 12 ground-truth objects.}
\footnotesize
\begin{tabular}{llccccccccccccc}
\toprule
& & \multicolumn{6}{c}{\textbf{No scene context}} & \multicolumn{6}{c}{\textbf{With scene context}} & \\
\cmidrule(lr){3-8} \cmidrule(lr){9-14}
\textbf{Scene} & \textbf{Model} & \textbf{F1} & \textbf{P} & \textbf{R} & \textbf{TP} & \textbf{FP} & \textbf{FN} & \textbf{F1} & \textbf{P} & \textbf{R} & \textbf{TP} & \textbf{FP} & \textbf{FN} & \textbf{$N$} \\
\midrule
\multirow{3}{*}{Office}
 & Pro        & $.223 \pm .171$ & $.438 \pm .309$ & $.157 \pm .136$ & 1.9 & 2.1 & 10.1 & $.295 \pm .132$ & $.629 \pm .202$ & $.203 \pm .117$ & 2.4 & 1.5 & 9.6 & 80 \\
 & Flash      & $\underline{.295 \pm .153}$ & $.480 \pm .212$ & $\underline{.220 \pm .133}$ & \underline{2.6} & 2.7 & \underline{9.4}  & $\underline{.385 \pm .122}$ & $.649 \pm .165$ & $\underline{.281 \pm .112}$ & \underline{3.4} & 1.8 & \underline{8.6} & 80 \\
 & Flash Lite & $.294 \pm .130$ & $\underline{.723 \pm .256}$ & $.191 \pm .096$ & 2.3 & \underline{0.9} & 9.7  & $.327 \pm .121$ & $\underline{.791 \pm .211}$ & $.214 \pm .097$ & 2.6 & \underline{0.8} & 9.4 & 80 \\
\midrule
\multirow{3}{*}{Kitchen}
 & Pro        & $\underline{.171 \pm .118}$ & $\underline{.404 \pm .279}$ & $\underline{.117 \pm .088}$ & \underline{1.4} & \underline{2.4} & \underline{10.6} & $.221 \pm .128$ & $.575 \pm .277$ & $.147 \pm .099$ & 1.8 & 1.4 & 10.2 & 50 \\
 & Flash      & $.157 \pm .073$ & $.302 \pm .185$ & $.113 \pm .060$ & \underline{1.4} & 3.7 & \underline{10.6} & $\underline{.316 \pm .131}$ & $\underline{.732 \pm .216}$ & $\underline{.212 \pm .106}$ & \underline{2.5} & \underline{1.0} & \underline{9.5} & 50 \\
 & Flash Lite & $.085 \pm .107$ & $.178 \pm .229$ & $.058 \pm .077$ & 0.7 & 2.8 & 11.3 & $.202 \pm .140$ & $.534 \pm .312$ & $.133 \pm .109$ & 1.6 & 1.3 & 10.4 & 50 \\
\midrule
\multirow{3}{*}{Campus}
 & Pro        & $\underline{.080 \pm .089}$ & $\underline{.208 \pm .233}$ & $.050 \pm .055$ & 0.6 & \underline{1.8} & 11.4 & $.278 \pm .080$ & $\underline{.860 \pm .172}$ & $.175 \pm .069$ & 2.1 & \underline{0.6} & 9.9  & 10 \\
 & Flash      & $.078 \pm .106$ & $.148 \pm .193$ & $\underline{.058 \pm .084}$ & \underline{0.7} & 3.4 & \underline{11.3} & $\underline{.347 \pm .115}$ & $.697 \pm .152$ & $\underline{.242 \pm .108}$ & \underline{2.9} & 1.3 & \underline{9.1}  & 10 \\
 & Flash Lite & $.059 \pm .060$ & $.107 \pm .113$ & $.042 \pm .042$ & 0.5 & 3.5 & 11.5 & $.222 \pm .096$ & $.502 \pm .187$ & $.158 \pm .102$ & 1.9 & 2.2 & 10.1 & 10 \\
\bottomrule
\end{tabular}
\label{tab:scene-model}
\end{table*}

\textbf{Best trials.} The strongest individual trial achieved $F1{=}.737$ (P1, brown eyes, office, Gemini Pro): 7 true positives with zero false positives, correctly identifying the monitor, secondary monitor, a cat image displayed on the monitor, the desk, coffee mug, headphones, and keyboard. The best kitchen trial reached $F1{=}.556$ (P1, brown eyes, Pro), detecting 5 objects: the faucet, sink, bowl, coffee cup, and window. The best campus trial reached $F1{=}.316$ (P2, brown eyes, Flash), detecting trees, a building, and a road sign.

\subsection{What Affects Capture Quality and Analysis?}
We examine how capture conditions affect performance. 
We varied phone distance, angle, motion, and lighting as primary conditions, and additionally analyzed physiological factors (eye color, contact lenses) across participants.

\begin{figure}[hbt!]
  \centering
  \includegraphics[width=.9\linewidth]{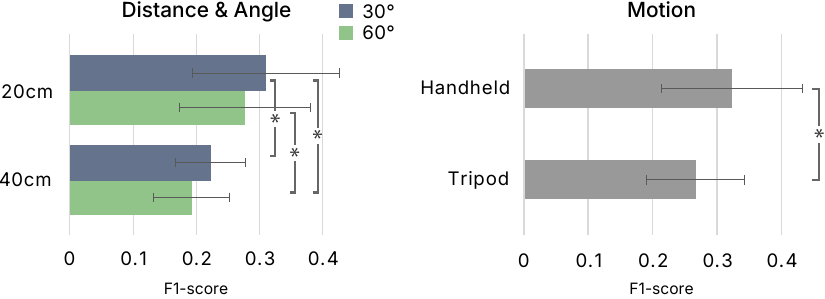}
  \caption{Capture conditions with significant effects on F1. Left: the distance $\times$ angle interaction, where 20\,cm outperformed 40\,cm at both angles. Right: handheld outperformed tripod. Error bars indicate standard deviation. (*: $p<.05$)}
  \label{fig:signif-plots}
\end{figure}

\textbf{Phone-to-eye distance.} Closer capture yielded significantly higher F1 score ($V{=}7$, $p < .05$): 20\,cm ($M_{F1}{=}.294$, $SD_{F1}{=}.101$) outperformed 40\,cm ($M_{F1}{=}.208$, $SD_{F1}{=}.051$).

\textbf{Phone-to-eye angle.} Phone angle alone (30\textdegree{} \vs 60\textdegree\ below eye level) did not significantly affect F1 ($V{=}15$, $p{=}.23$).

\textbf{Distance $\times$ angle interaction.} A Friedman test over the four conditions was significant ($\chi^2(3, N{=}10) = 8.04$, $p < .05$; \autoref{fig:signif-plots}, left), with 20\,cm/30\textdegree{} yielding the highest F1 ($M{=}.311$) and 40\,cm/60\textdegree{} the lowest ($M{=}.193$; $V{=}7$, $p < .05$). Post-hoc analyses indicate distance drove this interaction: 20\,cm outperformed 40\,cm at both 30\textdegree{} ($V{=}7$, $p < .05$) and 60\textdegree{} ($V{=}6$, $p < .05$), while angle was non-significant at either 20\,cm ($V{=}22$, $p{=}.63$) or 40\,cm ($V{=}13$, $p{=}.16$). 

\textbf{Motion.} Handheld capture ($M{=}.323$, $SD{=}.109$) significantly outperformed tripod ($M{=}.267$, $SD{=}.076$; $V{=}5$, $p < .05$; \autoref{fig:signif-plots}, right). Handheld capture allows natural head and arm movement, varying the reflected scene region across frames and giving the VLM more coverage of the surrounding objects.

\textbf{Lighting.} A Friedman test showed no significant effect of lighting ($\chi^2(2, N{=}10) = 1.08$, $p$=.58). Descriptively, medium ambient lighting ($M{=}.360$) outperformed high ($M{=}.310$) and low ($M{=}.270$).

\textbf{Eye color.} A Mann-Whitney U test showed no significant effect of eye color ($U{=}7.0$, $p{=}.31$), though dark eyes ($M{=}.262$, $SD{=}.053$) descriptively outperformed light eyes ($M{=}.199$, $SD{=}.058$). Dark-eyed participants also achieved higher ceilings across all three scenes (across all models): office $F1{=}.737$ (P1, brown, Pro) \vs $.632$ (P7, hazel, Flash); kitchen $.556$ (P1, brown, Pro) \vs $.353$ (P6, green, Pro); campus $.316$ (P2, brown, Flash) vs.\ $.143$ (P8, green, Pro), with 14 of 15 top-F1 trials coming from dark-eyed participants. This is optically intuitive: darker irises provide a higher-contrast background against which the specular corneal reflection is more visible.

\textbf{Contact lenses.} With only one contact-lens wearer per eye-color group, we report descriptive results only. Both fell below their group means: P9 (brown) $F1{=}.223$ \textit{vs.}\ $.272$; P5 (green) $.105$ \textit{vs.}\ $.222$. 

\subsection{What Do VLMs Detect, Misclassify, and Hallucinate?}
\label{sec:detect-hallucinate}
Here, we examine what Flash actually detects at the individual object level, and what it gets wrong.

\textbf{What is detectable.} Detection is driven by object size and contrast (\autoref{fig:tp-hallucination}a). In the office ($N{=}80$), the monitor was detected in 87.5\% of trials, the desk in 81.2\%, and the keyboard in 37.5\%. Even a cat image on the secondary monitor was identified in 11.2\% of trials. Other office objects appeared less frequently: secondary monitor and desk lamp (each $\sim$12\%), coffee mug and headphones (each 6.2\%), and mouse and notebook (each 5.0\%). In the kitchen ($N{=}50$), the window was detected in 100.0\% of trials, followed by a person (18.0\%), coffee cup (12.0\%), and faucet, sink, and bowl (each 2.0\%). On campus ($N{=}10$), trees were detected in 40\% of trials, with buildings, people, and a road sign each at 10.0\%.

During our review of the judge's scoring, we found that Flash also correctly identified objects outside our 12-object ground truth lists---\textit{e.g.,} office chair (33.8\% of trials) and laptop (18.8\%) in the office; wall art (16.0\%), bookshelf (12.0\%), and dish rack (2.0\%) in the kitchen; and sky (20.0\%) and sun (10.0\%) outdoors---suggesting VLMs can extract richer information than our scoring rubric captures.

\textbf{What is missed.} Many ground-truth objects were never detected: pen cup and eyeglasses (0/80 office trials); fridge, coffee machine, fork, cutting board, pot, and spatula (0/50 kitchen trials); and fountain, bench, grass, bushes, path, lamppost, and trash can (0/10 campus trials). Moreover, some had high miss rates, such as the secondary monitor (88.8\% missed), desk lamp (87.5\%), headphones (93.8\%), and coffee mug (93.8\%) in the office. For small objects (\eg, eyeglasses, fork), the limitation may be optical, yet others (\eg, headphones, coffee mug, fountain) were visually discernible to researchers, suggesting the bottleneck is partly VLM interpretation of this unfamiliar image domain (see \S\ref{sec:trajectory}).

\textbf{Most false positives are misclassifications, not
hallucinations.} An aggregate FP count does not separate what the optics revealed from what the VLM inferred, so we manually reviewed all 433 false-positive occurrences. The majority---293 (67.7\%)---were misclassifications, and this held consistently across scenes: 66.2\% in the office (141/213), 69.4\% in the kitchen (129/186), and 67.6\% on campus (23/34). This distinction matters for feasibility. A hallucination means the VLM reported something the cornea never captured. A misclassification means the cornea captured it and the VLM saw it, but named it wrong---which fine-tuning, better priors, or higher-resolution capture could address. The corneal signal may therefore carry more than the F1 scores alone convey.

\textbf{What is misclassified.} Misclassification patterns were scene-specific (\autoref{fig:tp-hallucination}b). In the kitchen ($N{=}50$), Flash projected office objects onto the scene: ``monitor'' appeared in 86.0\% of trials and ``desk'' in 82.0\%. Our audit traced both to real optical evidence---``monitor'' typically corresponded to the bright interior of the open fridge or an illuminated window (43 occurrences), and ``desk'' to the brown cutting board and countertop (41). In the office ($N{=}80$), ``window'' was the dominant misclassification (87.5\% of trials, 70 occurrences), triggered by the bright monitor screen, while ``ceiling light'' and ``light source'' (39 combined) corresponded to the desk lamp. On campus ($N{=}10$), ``window'' (80.0\%) corresponded to building facades and ``desk'' (50.0\%) to a wooden bench. The common thread is that at this resolution many pixel patches are visually ambiguous---for example, a bright rectangular region could equally be a monitor, window, open fridge, or desk lamp.

\textbf{What is hallucinated.} The remaining 140 false positives (32.3\%) had no identifiable optical evidence (\autoref{fig:tp-hallucination}c). In the kitchen, the projection of office objects extended to ``keyboard'' (23 occurrences), ``lamp'' (9), ``mouse'' (9), and ``laptop'' (7). In the office, ``person'' was the most frequent hallucination (27 occurrences) despite no person being present, followed by ``blinds'' (16) and ``trees'' (8). On campus, hallucinations were sparse (11 occurrences) and led by ``monitor'' (4). Notably, both the misclassifications and hallucinations in the kitchen were predominantly office objects, suggesting a cascade: an office label applied to the reflection of a kitchen object shifts the scene guess, biasing future VLM calls to report office objects that were never there. We discuss this further in \S\ref{sec:trajectory}.

\begin{figure*}[t]
  \centering
  \includegraphics[width=\linewidth]{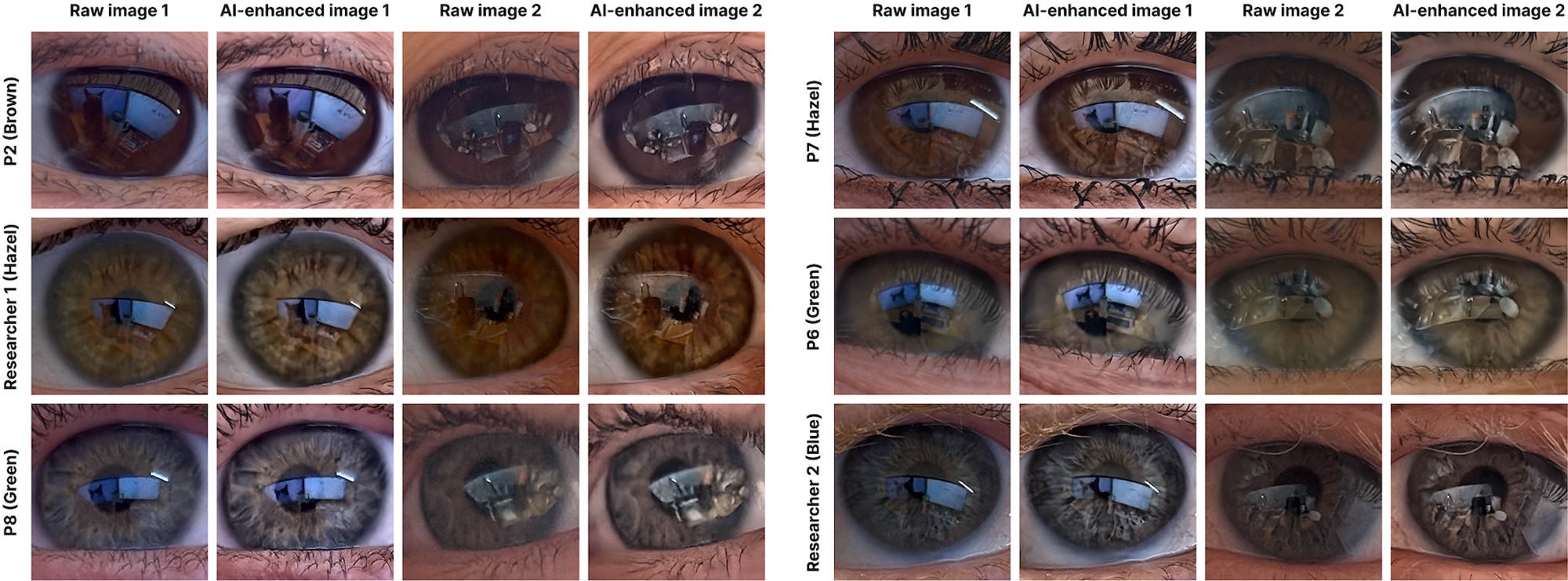}
  \caption{Participants' and researchers' raw–enhanced corneal pairs from darker (top left) to lighter (bottom right) eye colors}
  \label{fig:participant-eyes}
\vspace{-0.25cm}
\end{figure*}

\subsection{Can Minimal Scene Context Reduce Errors?}
\label{sec:context-ablation}
Many objects were optically visible in the corneal reflection, yet
VLMs failed to name them correctly. We therefore asked whether adding
context from other sensors could close that gap. Phones already have
access to other context sources, such as GPS, time, motion, and
ambient light. To simulate fusing these with corneal reflections, we
prepended a single short sentence naming the scene (\eg, ``The user is
in a kitchen'') and re-ran the full evaluation (\S\ref{sec:analysis}).

\begin{figure*}[hbt!]
  \centering
  \includegraphics[width=\linewidth]{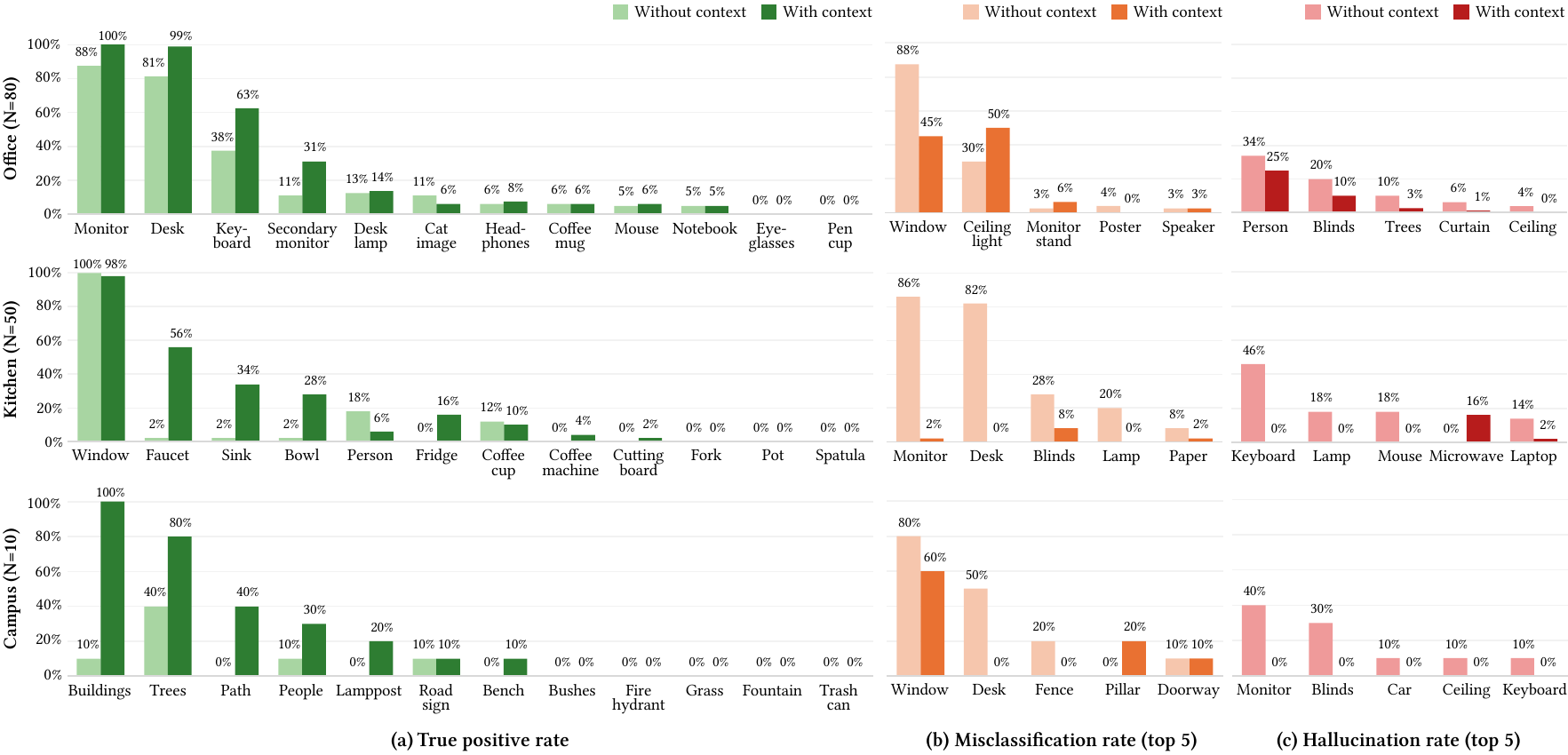}
  \vspace{-12px}
  \caption{Per-object rates by scene, without and with a minimal scene prior (Flash): (a) true positive rate against the 12 ground-truth objects, (b) top-5 misclassification rate, and (c) top-5 hallucination rate. Additional context improved detection of objects the VLM could already see but misidentified, and wrong-scene errors (\eg, ``monitor'' and ``desk'' in the kitchen) largely vanished.}
  \vspace{-8px}
  \label{fig:tp-hallucination}
\end{figure*}

\textbf{Scene context substantially improved performance.} Flash F1 rose from $.230$ to $.358$ overall, with precision nearly doubling ($.393 \rightarrow .682$)---suggesting that even a minimal scene prior was enough to reduce wrong guesses. True positive count increased from 286 to 426 (+49.0\%), while false positives dropped from 433 to 210 ($-$51.5\%). The improvement was most dramatic where performance had been worst: campus F1 rose from $.078$ to $.347$, with building detection jumping from 10.0\% to 100.0\%, trees from 40.0\% to 80.0\%, and previously invisible objects now detected (path: 0.0\% $\rightarrow$ 40.0\%, people: 10.0\% $\rightarrow$ 30.0\%, lamppost: 0.0\% $\rightarrow$ 20.0\%, bench: 0.0\% $\rightarrow$ 10.0\%). Kitchen F1 improved from $.157$ to $.316$, with faucet rising from 2.0\% to 56.0\%, sink from 2.0\% to 34.0\%, bowl from 2.0\% to 28.0\%, and fridge from 0\% to 16.0\% of trials. Office F1 improved from $.295$ to $.385$, with the monitor reaching 100.0\% detection, the desk reaching 98.8\%, and the keyboard rising from 37.5\% to 62.5\% (\autoref{fig:tp-hallucination}). Best-case performance also improved: the strongest kitchen trial with context detected 7 objects ($F1{=}.556 \rightarrow .700$, 1 false positive)---including the fridge and cutting board, both undetectable without context---compared to 5 objects without. The campus ceiling rose from 3 to 6 detected objects ($F1{=}.316 \rightarrow .571$), now including path, bench, and person. This suggests corneal reflections become more powerful when fused with other context a phone already collects.

\textbf{Scene context reduced both misclassification and hallucination.} Re-auditing false positives with scene prior, misclassifications fell from 293 to 150 ($-$48.8\%) and hallucinations from 140 to 60 ($-$57.1\%). In the kitchen scenario, office-object errors largely vanished: ``monitor'' dropped from 86.0\% of trials to 2\% and ``desk'' from 82.0\% to 0\%, while the hallucinated ``keyboard'' fell from 46.0\% to 0\%. The remaining kitchen errors stayed within the correct scene: misclassifications such as ``small appliance'' and ``overhead light,'' and ``microwave'' (8 occurrences) as the top hallucination. Errors shifted from impossible to plausible. That said, the model also began using more generic labels (\eg, ``small appliance''), despite prompt instructions to be specific.

\textbf{Scene context reduced the influence of capture conditions.} With context, the distance ($p<.05 \rightarrow p{=}.160$) and motion ($p<.05 \rightarrow p{=}.375$) effects were no longer significant. This suggests a minimal scene prior reduced sensitivity to suboptimal capture.

\textbf{What scene context did not fix.} Despite improvements, pen cup, eyeglasses, fork, pot, spatula, fountain, grass, fire hydrant, bushes, and trash can remained at 0.0\% detection with or without context. These objects are too small or too low-contrast to produce a resolvable reflection. This suggests the added scene context was not simply forcing VLMs to make plausible guesses, since if that were the case, obvious kitchen and outdoor objects like the fork and grass would have appeared. A scene prior appears to help VLMs resolve visual ambiguity in low-resolution reflection patches, largely without biasing them to invent what the cornea never captured.

\subsection{Participant Perceptions}
We collected 5-point Likert ratings (1: very uncomfortable, 5: very comfortable) and open-ended responses before and after the study, probing comfort and privacy concerns around corneal reflection.

\textbf{Privacy perceptions.} Before the study, participants were slightly uncomfortable with corneal reflection ($M$=2.7, $SD$=0.95), with concerns about surveillance (P1, P7) and not having autonomy over what is captured (P9, P10). Comfort increased in a hypothetical where data stayed on-device ($M$=3.9, $SD$=0.99), though some remained skeptical: P9 worried ``\textit{what if other people still found ways to have my corneal reflection?}'' and P1 asked ``\textit{what if my phone gets hacked?}'' The main post-study concern was \textit{stealthiness}---that corneal capture is invisible to bystanders, unlike a visible camera. P1 felt ``\textit{less sure if a person is capturing or not},'' and P7 noted ``\textit{AI is getting better\ldots so they might find out everything about me}.''

\textbf{Reactions to seeing their own corneal images.} After the capture trials but before seeing their own eye images, participants were moderately comfortable ($M$=3.3, $SD$=0.82), with six participants assuming the images would not reveal much. After seeing their own corneal reflections, comfort dipped slightly ($M$=3.0, $SD$=1.25) but reactions diverged. Five participants (P1, P2, P3, P6, P7) found images clearer than expected, which often decreased comfort: P1 dropped from 4 to 2 (``\textit{it revealed more than I thought\ldots a bit scary}''), and P3 noted that seeing ``\textit{specific components, like a handle of a cup}'' made them ``\textit{less confident with sharing}.'' Five others (P4, P5, P8, P9, P10) found images blurrier, which sometimes increased comfort: P5 rose from 4 to 5, and P4 noted ``\textit{they still have relatively low quality, which I actually like}.'' Despite these concerns, interest in trying a future version remained high ($M$=4.1, $SD$=0.74). Six participants saw value in a camera that can capture surroundings during ordinary phone use, when neither front nor rear camera faces the world.

\textbf{Corneal \vs RGB photo capture.} When comparing corneal reflection to a rear-camera view ($M$=2.5, $SD$=1.08; 1: prefer rear camera, 5: prefer corneal image), participants surfaced tension between convenience, control, and presence. Some valued that corneal capture lets users stay present in the moment rather than experiencing life through a phone screen---P9 noted ``\textit{if I'm hiking or in a safari, looking at the world is quite important},'' and P6 imagined ``\textit{a report of moments in your life}'' without being on the phone. Others preferred the intentionality of a rear camera: P6, who saw merit in both, valued ``\textit{the feeling of controlling the camera myself}'', and P7 wanted ``\textit{more choice over what is being captured}.'' Participants envisioned various applications, including pedestrian safety (P1, P9), screen-free photo capture (P6, P9), accessibility for blind or low-vision users (P9), attention monitoring (P2), and grocery assistance (P5, P7). We implemented two of these---pedestrian safety and attention monitoring---as demonstration scenarios (\S\ref{sec:proof-by-demonstration}).

\begin{figure*}[hbt!]
  \centering
  \includegraphics[width=\linewidth]{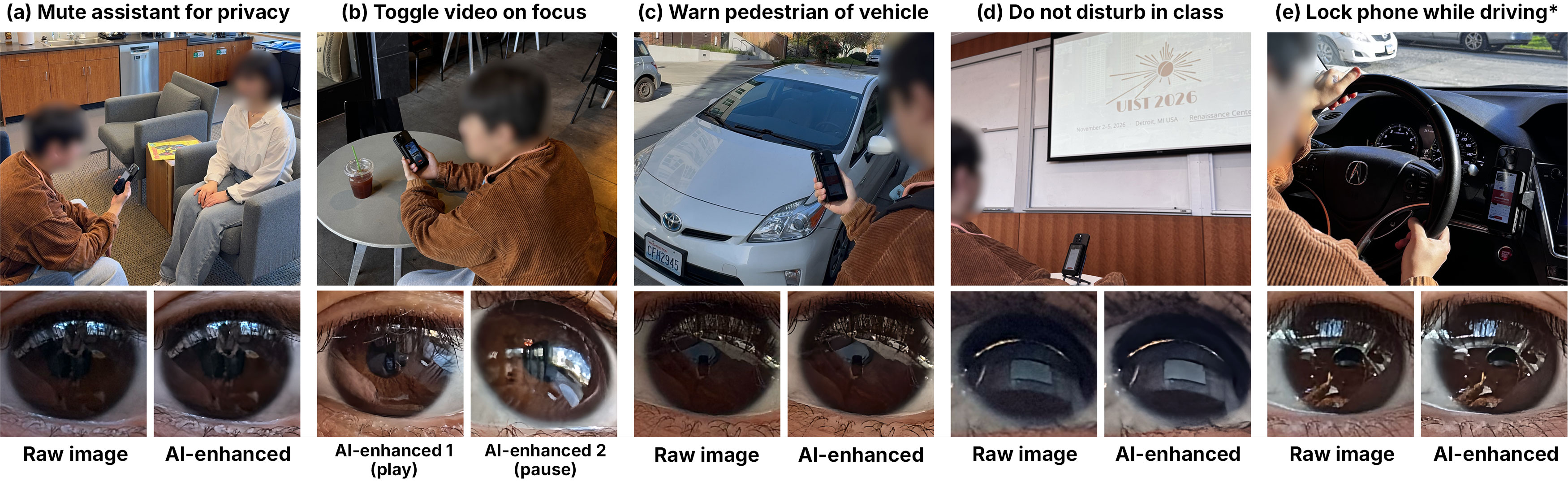}
  \vspace{-16px}
  \caption{Proof-by-demonstration scenarios. Each column shows a real-world scenario (top) and the corresponding best raw and AI-enhanced corneal images (bottom). All five interactions were successfully triggered by objects detected in live corneal reflections. (*driver detection required a scene-context prior)}
  \vspace{-10px}
  \label{fig:pbd}
\end{figure*}

\section{Proof by Demonstration}
\label{sec:proof-by-demonstration}
To explore the real-world potential of corneal reflections as a context
source, we built five interaction scenarios on \sysname's live pipeline
(\autoref{fig:pbd}). We selected these scenarios because each requires knowing something about the surroundings, yet asking users to aim a camera would be impractical. In many, no world-facing view is available at all: a phone held below eye level points its rear camera at the ground, and a phone mounted in a car points it at the dashboard. In others, we used the phone as a proxy for laptops and webcams, which have no world-facing camera.

The scenarios are: (1) a \textit{Social Presence Guard} that detects nearby people to pause a voice assistant, (2) an \textit{Attention Tracker} that detects screen drift to auto-play or pause video, (3) a \textit{Pedestrian Safety Alert} that warns phone-distracted users of approaching vehicles, (4) a \textit{Courtesy Switch} that recognizes a presentation surface to activate Do Not Disturb, and (5) a \textit{Driver Detector} that identifies a steering wheel to restrict phone use. In all of these, corneal reflection is the only reasonable view of the scene these devices have.

All five triggered correctly: \sysname detected the relevant objects from live corneal reflections and fired the intended action. VLMs identified ``person'' to pause the assistant; ``phone'' to play and ``window'' to pause video; ``vehicle'' and ``road'' to warn the pedestrian; ``whiteboard'' and ``screen'' to trigger Do Not Disturb; and ``steering wheel'' and ``dashboard'' to infer driving. Four required no scene prior; only driver detection needed one (\ie, ``the user is in a car''). Detection latency varied, requiring 2--4 VLM returns ($\sim$23 seconds) before a stable scene state emerged (see supplementary video).

More interactions become possible once a device with only a user-facing camera can sense what is in front of the user. On a phone, a user reading a grocery list could have the aisle recognized, surfacing relevant coupons without lifting and pointing the rear camera. Beyond phones, the Social Presence Guard could run on an office webcam, detecting someone approaching from behind the screen and dimming the display to prevent shoulder surfing. The Courtesy Switch could run on a student's laptop in a classroom. On smartglasses, an inward-facing camera could supply both attention tracking and scene context, reducing the sensing hardware such devices require. While challenges remain, corneal-reflection sensing can already support meaningful context-driven interactions.

\section{Discussion}
\label{sec:discussion}
As an initial exploration of corneal-reflection sensing with smartphone-grade optics, our primary goal was to establish whether scene understanding from this low-resolution signal is feasible, and what remains open. The results are encouraging: off-the-shelf VLMs classified indoor \vs outdoor at 96.4\% and detected real objects across all three scenes, with the VLM more often misclassifying what the cornea captured than hallucinating what it did not. We discuss where corneal sensing is headed, the privacy questions it raises, and the limitations of this work.

\subsection{Where Corneal Sensing Is Headed}
\label{sec:trajectory}
We begin by enumerating what corneal sensing can already achieve, before turning to its limitations and trajectory.

\textbf{What the cornea already reveals.} Our results suggest corneal sensing can support coarse context tasks---distinguishing indoors from outdoors, detecting bigger objects like a monitor, desk, faucet, or steering wheel, and sensing nearby people---while collapsing the two-camera requirement of prior systems~\cite{lander2017heyebrid, mayer2020worldgaze} into one. More fundamentally, these findings \textit{reframe} the eye in HCI: beyond an input channel for gaze and attention~\cite{jacob2003eye, majaranta2014eye}, the cornea can act as a \textit{passive scene camera}. Yet current performance is almost certainly a lower bound. VLMs see far more than they name: 67.7\% of false positives were misclassifications rather than hallucinations, one sentence of scene prior raised true positives 49\% while halving false positives, and Flash produced 346 correct detections outside our ground truth. Several paths could raise this floor, such as using the full RGB face frame as scene context before cropping, video or burst super-resolution across consecutive corneal frames~\cite{nitschke2012super, alzayer2024seeingtheworld, wronski2019handheld} when front cameras catch up, and fine-tuning image restoration models and VLMs for this domain they presumably never encountered. Smartglasses, with eye-facing cameras at close range~\cite{bulling2016eyewear}, could also use corneal reflections given that closer capture improves performance. Moreover, with the phone propped $\sim$115\,cm from the eye, \sysname still detected the window in every kitchen trial, suggesting future smartphone cameras could better support handheld use. Longer-term, the gaze coupling is intriguing; since the pupil is darker than the iris, the reflection center is roughly aligned with gaze~\cite{takemura2014estimation}, potentially enabling joint scene and attention inference.

\textbf{Challenges of continuous VLM inference.} To build scene understanding over time, \sysname chains VLM calls across successive batches, with each call receiving the previous state as JSON rather than the images behind it. This was among our most difficult challenges, and two key problems still remain. First, \textit{error cascading}: a misclassified pixel patch (\eg, a window reflection in the kitchen labeled ``monitor'') biases the scene guess (\eg, toward
``office'') which increases wrong-scene predictions (\eg, ``desk,''
``keyboard'') and biases future calls. Because only the JSON state
carries forward, a wrong label cannot be re-checked against the pixels
that produced it. Second, the pipeline lacks \textit{object
persistence}: if a monitor leaves and re-enters the frame, \sysname
cannot tell whether it is the same one or a new detection. More
broadly, chained inference must balance being too destructive
(overwriting correct detections) and too trusting (accumulating
errors)---a tension that affects any live VLM system, but amplified by
corneal imagery's low resolution and per-frame confidence. Stricter frame gating before inference, or passing some images alongside textual state, could reduce this cascading. This may also partly explain the gap between average and best-case performance, where the strongest office trial reached $F1{=}.737$ against a scene mean of $.295$. Relatedly, in manually auditing the corneal images, two researchers could often identify objects the VLM did not name; see \autoref{fig:participant-eyes} and Appendix~\ref{sec:researcher-eyes} for examples.

\textbf{Hard limits.} Even with video-rate capture, better cameras, and fine-tuned models, some constraints will remain. Optically, the cornea is a convex mirror roughly 7.8\,mm in radius~\cite{mashige2013review}, compressing a wide-angle view of the scene into a few millimeters where small objects will only take up a few pixels. Many objects at 0\% detection (\eg, pen cup, fork, spatula) were small or low-contrast. Biologically, eye color, contact lenses, and eyelash occlusion introduce per-frame variance that temporal aggregation cannot resolve (\eg, 14/15 top-F1 trials came from dark-eyed participants). Practically, the eye has to be in frame, which users will not consciously maintain. Direct sunlight can also wash out reflections. \sysname is ultimately bounded by the cornea's ability to reflect the scene. But corneal sensing need not match a world-facing camera, as even coarse context already enables useful interactions.

\subsection{Privacy as Both Risk and Opportunity}
\label{sec:privacy-discussion}
Corneal-reflection sensing raises privacy questions that differ from those of conventional cameras. We discuss three dimensions: stealthiness of capture, potential platform-level mitigations, and tension around improving this technology.

\textbf{Stealthiness and consent.} Unlike a rear-facing camera, corneal capture raises a two-sided transparency problem: users cannot easily verify what is inferred from their eyes, and bystanders cannot tell they are being captured at all. For users, future systems could incorporate a ``what does my eye reveal?'' preview and a visible capture indicator, following the transparency patterns modern OSes use for camera and location access. For bystanders, the problem is harder: corneal data is a side effect of any face-facing photo and video, making capture difficult to signal---though corneal reflections could be removed entirely (see below). Despite these concerns, participants rated interest in a future version at 4.1/5, and proposed applications we went on to prototype (pedestrian safety, attention monitoring), suggesting the interaction value is compelling enough to try to navigate the privacy tension.

\textbf{Platform-level mitigations.} Rather than treating corneal reflections purely as a threat, we see an opportunity for proactive platform-level protection. Since smartphone cameras already apply computational photography to each image~\cite{wronski2019handheld, hasinoff2016burst}, adding corneal-region inpainting---analogous to red-eye reduction~\cite{gaubatz2002automatic}---could mitigate this side channel before images reach applications. This is urgent, as selfies are already high-resolution, likely containing corneal data that could be analyzed retroactively. In fact, in 2019 a stalker located a Japanese pop idol by enlarging corneal reflections in her public selfies, identifying her local train station~\cite{guardian2019stalker}. Fortunately, on-device protection is increasingly feasible, as our restoration models are lightweight and on-device VLM inference is rapidly maturing~\cite{fastvlm2025}, ensuring corneal data need never leave the phone. Our participants' comfort ratings support this, rising from $M$=2.7 to 3.9 in a hypothetical where eye data remains on-device.

\textbf{The fine-tuning paradox.} An obvious next step is training VLMs on corneal imagery, yet despite collecting over 10,000 images, we chose not to given privacy concerns. Our IRB did not permit dataset or model release, and many participants opted out of image inclusion. This tension is inherent: advancing corneal sensing requires data that ethical practices restrict. It cuts deeper---should we \textit{want} this capability to improve? Better corneal understanding enables compelling interactions from a single camera, but also more potent surveillance~\cite{backes2008compromisingreflections, backes2009tempestinateapot}. We believe the benefits warrant continued research, provided privacy is handled carefully. One path is synthetic corneal data, rendering scenes onto eye models~\cite{kim2019nvgaze, du2024cred} to avoid real-eye data collection. Until such datasets exist, our results show what off-the-shelf VLMs achieve with no domain adaptation---itself a meaningful zero-shot benchmark for future work.

\subsection{Limitations and Future Work}
As the first exploration of VLM-based corneal scene understanding, our study prioritized breadth---testing multiple scenes, conditions, and models---over depth in any single dimension. Our sample ($N$=10) is small; a larger study is needed to draw more robust conclusions about physiological factors such as eye color effects. \sysname also captures with the rear camera as a stand-in for higher-resolution front sensors (\S\ref{sec:client}), and we tested only three scenes---environments such as moving vehicles or crowded public spaces may behave differently. Finally, we excluded glasses wearers, a substantial fraction of the population, and evaluated only the Gemini model family. A formal comparison of human and VLM performance on the same crops would help separate what the cornea carries from what current models can name.

\section{Conclusion}
In this work, we explored whether corneal reflections captured with smartphone-grade optics can serve as a source of scene context, and whether off-the-shelf VLMs can read them. We built \textit{\sysname}, an end-to-end pipeline pairing iris-targeted capture with server-side restoration and progressive VLM analysis. Across 10,635 corneal images from ten participants, VLMs classified indoor \vs outdoor reliably and detected objects in every scene, and most errors were misclassifications of real optical evidence rather than hallucinations. Because we did not fine-tune, these results serve as a zero-shot benchmark for future work. We hope this work motivates the HCI and AI communities to explore this untapped sensing channel---with the privacy care this intimate signal demands.


\bibliographystyle{ACM-Reference-Format}
\bibliography{paper}

\clearpage
\onecolumn
\appendix
\section{Full Eye Collage}
\label{sec:researcher-eyes}
\vspace{8px}
\begin{minipage}{\textwidth}
  \centering
  \includegraphics[width=\textwidth,height=.95\textheight,keepaspectratio]{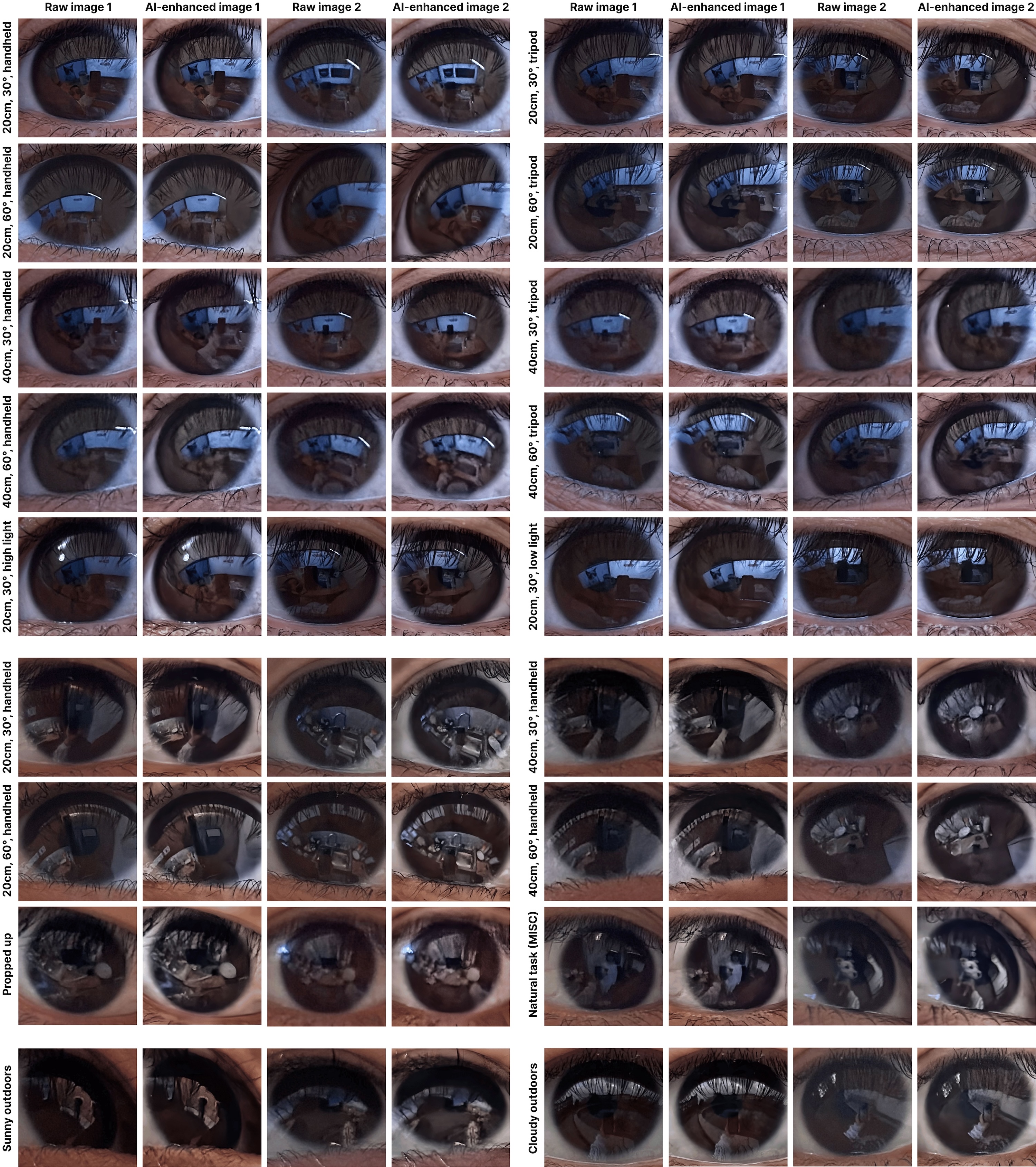}
  \captionof{figure}{A researcher's (dark brown eyes) corneal reflection images across all study trials}
  \label{fig:researcher-eyes}
\end{minipage}

\section{VLM Scene Analysis Prompt}
\label{sec:vlm-prompt}
\begin{minted}[breaklines,frame=single,framesep=1mm,baselinestretch=0.9,fontsize=\small]{text}
BASE_PROMPT = """
You are analyzing corneal reflection images - photographs of a human eye where the curved corneal surface acts as a convex mirror, capturing a wide-angle reflection of the surrounding environment. Think of it like a security mirror or a chrome ball reflection, but smaller and on an eye.

Visual characteristics to expect:
- The reflection appears in the pupil/iris region
- Images are warped due to the cornea's curved geometry
- Details may be low-resolution, noisy, or slightly blurred
- Lighting sources often appear as bright spots
- Shapes are distorted but preserve relative spatial relationships

The corneal reflection image was captured by a smartphone camera. You will likely see the smartphone (e.g., a dark rectangle with camera lenses) and the user's hand/arm or a tripod device holding it. Ignore the smartphone - it is the capture device, not part of the scene. Focus on describing everything else in the reflection.

You are receiving multiple image pairs. Each pair consists of:
- A RAW corneal crop (lower but decent resolution, unprocessed - preserves true highlight and edge detail)
- Its AI-ENHANCED version (4x resolution via Real-ESRGAN, improved contrast - may reveal more detail but could have minor artifacts)

Both images in a pair show the same reflection from the same moment. Use whichever (or both) provides clearer detail for each object. The raw version is more trustworthy for object details, while the enhanced version may reveal more dark or low-contrast objects/areas.

This is a multi-call pipeline: your JSON response will be passed as context to the next VLM call processing the next batch of images. Errors, hallucinations, or noise in your output will propagate and mislead future calls. Accuracy matters more than completeness - a short, correct list is far better than a long, noisy one. Image quality varies significantly between batches - many frames will be blurry or uninformative, with occasional clear frames that reveal real detail. Do not force observations from poor-quality images.

Within each batch, image quality will vary. Focus your analysis on the clearest pairs. Blurry or motion-affected pairs should be disregarded rather than interpreted - forcing meaning from noise is worse than skipping a frame.

Object naming rules:
- Be concise and specific (e.g., "giraffe", not "tall figure with four legs and a tail, which could be an elephant or giraffe")
- Before adding an object, check if it already exists under a synonym or different name (e.g., "sneakers" and "running shoes" are the same object). Keep the more specific name. Do NOT add duplicates.
- Partially visible or blurry objects can be identified through educated guesses using surrounding context. Pure guesses from ambiguous blobs with no contextual support should be omitted entirely.
- Do not let any examples in these instructions bias your observations. These examples are purely illustrative. Identify objects based solely on visual evidence in the corneal reflection.
"""

FIRST_CALL_CONTEXT = """
This is the FIRST batch of corneal reflection images. No previous scene analysis exists yet.

STEP 1 - OBSERVE: Examine these images and describe what you can see in the reflections. Identify the scene type, a specific scene name, and any distinct objects.

STEP 2 - RECORD: Build the initial scene understanding. For each object, assign a confidence level:
- "confirmed": Clearly identifiable - you can see its shape, color, or distinctive features
- "tentative": Partially visible, blurry, or obscured, but context and educated reasoning suggest what it is

If the images are too blurry to identify anything with reasonable confidence, it is acceptable to output a minimal response with just scene_type and scene_name. Do not pad the object lists with guesses from ambiguous blobs."""

SUBSEQUENT_CALL_CONTEXT = """
STEP 1 - OBSERVE: Examine these new images and describe what you can see in the reflections on their own - the scene type, a specific scene name, any objects, and spatial layout. Do this BEFORE looking at the previous scene understanding below.

STEP 2 - RECONCILE: Now, merge your fresh observations with the existing scene understanding.

CURRENT SCENE UNDERSTANDING (built from {window_count} previous batch(es)):
{master_json}

Merging rules:
- If your independent observation AGREES with the current understanding, reinforce it. Promote "tentative" objects to "confirmed" when you can clearly see their shape, color, or distinctive features in the current images - not simply because they appeared in a previous batch.
- If your independent observation CONTRADICTS the scene_type or scene_name, and the current images provide clear visual evidence for the correction, update it. Remove objects that no longer make sense in the corrected scene. Do not override an established scene label based on ambiguous or blurry images alone.
- ADD new objects with appropriate confidence. "confirmed" = clearly identifiable from the image evidence alone (distinct shape, color, or features). "tentative" = educated guess from partial, blurry, or obscured views where context suggests what it is. Check for duplicates and synonyms before adding.
- "tentative" objects may be removed if they seem inconsistent with accumulated evidence. "confirmed" objects should rarely be removed.
- If this batch of images is too blurry or low-quality to confidently identify anything new, return the previous understanding UNCHANGED. Do not add objects from ambiguous blobs just to contribute something.

IMPORTANT - confirmation bias: You are susceptible to anchoring on what is already in the master. An object appearing in the previous master does NOT make it more likely to be real. For each tentative object you are about to keep or promote, ask yourself: if this object were NOT already in the master, would the current images alone give me enough reason to add it? If no, do not promote it. Tentative objects should only be promoted to confirmed when you can independently identify them in the current batch - not because they have "been in the list for a while."

CRITICAL: The corneal reflection only captures a PORTION of the surrounding scene at any given time. As the eye moves, different objects enter and leave the reflection. The ABSENCE of an object in the current batch does NOT mean it was wrong - it simply may not be in the reflection right now. NEVER remove an object just because you don't see it in these specific images."""

OUTPUT_FORMAT = """
Respond with ONLY valid JSON in this exact format:

{
    "scene_type": "<indoor or outdoor>",
    "scene_name": "<specific scene name, e.g., bedroom, office, cafe, kitchen, courtyard, garden, living room, park, mall, etc.>",
    "confirmed_objects": ["<clearly identifiable - distinct shape, color, or features>"],
    "tentative_objects": ["<educated guess from partial or blurry view>"],
    "scene_description": "<One detailed sentence describing the overall scene, visible objects, and spatial arrangement>"
}"""
\end{minted}
\section{LLM Judge Prompt}
\label{sec:llm-prompt}
\begin{minted}[breaklines,frame=single,framesep=1mm,baselinestretch=0.9,fontsize=\small]{text}
JUDGE_PROMPT = """
You are evaluating a Vision Language Model's scene analysis accuracy based on corneal reflections.

## Ground Truth
- scene_type: "{gt_scene_type}"
- scene_name: "{gt_scene_name}" (also acceptable: {gt_aliases})
- objects present in the scene: {gt_objects}

## Explicit Object Aliases (ALWAYS accept these matches)
{object_aliases}

## VLM Response
- scene_type: "{vlm_scene_type}"
- scene_name: "{vlm_scene_name}"
- confirmed_objects: {vlm_confirmed}
- tentative_objects: {vlm_tentative}

## Your Task
1. Check if scene_type matches (indoor/outdoor)
2. Check if scene_name is semantically equivalent to ground truth or any alias
3. For EACH VLM-detected object (both confirmed and tentative):
   - FIRST check explicit aliases above - if VLM output matches any alias, ACCEPT it
   - If NOT in explicit aliases, use your best judgment:
     - Accept clear synonyms at similar specificity (e.g., "display" -> "monitor")
     - Accept functional equivalents only if specific enough (e.g., "computer screen" -> "monitor")
     - Reject if the term could match multiple ground truth objects (e.g., "light" when both "desk lamp" and "ceiling light" exist)
     - Reject overly vague/generic terms: "device", "furniture", "electronics", "object", "thing", "bright rectangle", "dark shape"
     - When genuinely uncertain, reject - we reward specificity
4. Identify which ground truth objects were NOT detected by the VLM (in either confirmed or tentative)

IMPORTANT: Evaluate confirmed and tentative objects TOGETHER as a single pool of detected objects. Both count equally for matching purposes.

Respond with ONLY valid JSON in this exact format:
{{
    "scene_type_correct": <true or false>,
    "scene_name_correct": <true or false>,
    "true_positives": {{"<vlm_object>": "<matched_gt_object>", ...}},
    "false_positives": ["<vlm_object_not_in_gt>", ...],
    "false_negatives": ["<gt_object_not_detected>", ...]
}}
"""
\end{minted}

\begin{minted}[breaklines,frame=single,framesep=1mm,baselinestretch=0.9,fontsize=\small]{text}
{
   "office": {
       "scene_type": "indoor",
       "scene_name": "office",
       "scene_name_aliases": ["work desk", "home office", "workspace", "study", "desk area", "workstation", "computer desk"],
       "objects": [
           "monitor",
           "secondary monitor",
           "mouse",
           "keyboard",
           "notebook",
           "pen cup",
           "desk lamp",
           "coffee mug",
           "eyeglasses",
           "headphones",
           "cat image",
           "desk"
       ],
       "object_aliases": {
           "monitor": ["display", "screen", "computer screen", "computer monitor", "primary monitor", "main monitor"],
           "secondary monitor": ["second monitor", "second display", "second screen", "dual monitor", "side monitor"],
           "mouse": ["computer mouse"],
           "keyboard": ["computer keyboard"],
           "notebook": ["stack of books", "books", "book", "notepad"],
           "pen cup": ["pencil cup", "pen holder", "pencil holder", "pen container", "stationery holder", "cup of pens"],
           "desk lamp": ["table lamp", "lamp", "desk light"],
           "coffee mug": ["mug", "cup", "coffee cup"],
           "eyeglasses": ["glasses", "spectacles", "reading glasses"],
           "headphones": ["headset", "earphones", "over-ear headphones"],
           "cat image": ["cat", "kitten"],
           "desk": ["table", "work desk", "computer desk", "workstation"]
       }
   },
   "kitchen": {
       "scene_type": "indoor",
       "scene_name": "kitchen",
       "scene_name_aliases": ["kitchenette", "cooking area", "kitchen counter", "kitchen area"],
       "objects": [
           "fridge",
           "sink",
           "faucet",
           "coffee machine",
           "coffee cup",
           "bowl",
           "fork",
           "cutting board",
           "pot",
           "spatula",
           "person",
           "window"
       ],
       "object_aliases": {
           "fridge": ["refrigerator", "freezer"],
           "sink": ["kitchen sink"],
           "faucet": ["kitchen faucet", "tap", "water tap"],
           "coffee machine": ["coffee maker", "espresso machine", "coffee brewer"],
           "coffee cup": ["mug", "cup", "coffee mug"],
           "bowl": ["mixing bowl", "cereal bowl", "white bowl"],
           "fork": ["utensil", "silverware"],
           "cutting board": ["chopping board"],
           "pot": ["cooking pot", "saucepan", "pan"],
           "spatula": ["cooking utensil", "flipper"],
           "person": ["human", "someone", "figure", "man", "woman", "people"],
           "window": ["kitchen window", "windows"]
       }
   },
   "outdoor": {
       "scene_type": "outdoor",
       "scene_name": "campus",
       "scene_name_aliases": ["university quad", "quad", "campus quad", "courtyard", "university courtyard", "campus courtyard", "plaza", "college campus"],
       "objects": [
           "fountain",
           "bench",
           "trees",
           "grass",
           "fire hydrant",
           "bushes",
           "people",
           "path",
           "lamppost",
           "trash can",
           "road sign",
           "buildings"
       ],
       "object_aliases": {
           "fountain": ["water fountain"],
           "bench": ["stone bench", "wooden bench", "seat"],
           "trees": ["tree", "large tree", "bare tree"],
           "grass": ["lawn", "green grass", "grassy area", "turf"],
           "fire hydrant": ["hydrant"],
           "bushes": ["bush", "shrubs", "shrub"],
           "people": ["person", "pedestrians", "pedestrian", "human", "humans", "figures", "passersby", "woman", "man"],
           "path": ["walkway", "sidewalk", "pathway", "walking path", "footpath", "trail", "pavement"],
           "lamppost": ["lampposts", "street light", "street lamp", "light pole", "street lights", "lamp post", "light post"],
           "trash can": ["garbage can", "waste bin", "bin"],
           "road sign": ["sign", "street sign", "signage"],
           "buildings": ["building", "structure", "structures"]
       }
   }
}
\end{minted}

\end{document}